\documentclass[12pt]{iopart}
\usepackage{graphicx}
\begin{document}

\title[TDDFT approach to nonlinear particle-solid interactions]
{Time-dependent density-functional theory approach to
nonlinear particle-solid interactions in comparison with
scattering theory}

\author{V U Nazarov\dag\S, J M Pitarke\ddag$\|$, C S Kim\dag\ and
Y Takada\S}

\address{\dag\ Department of Physics and
 Institute for Condensed Matter Theory,
 Chonnam National University,
 Gwangju 500-757, Korea}

\address {\S\ Institute for Solid State Physics, University of
Tokyo, Kashiwa, Chiba 277-8581, Japan}

\address {\ddag\ Materia Kondentsatuaren Fisika Saila, Zientzi Fakultatea, Euskal
Herriko Unibertsitatea, 644 Posta Kutxatila, E-48080 Bilbo, Basque Country, Spain}

\address {$\|$\ Donostia International Physics Center and Centro
Mixto CSIC-UPV/EHU, Donostia, Basque Country, Spain}

\ead{nazarov@boltzmann.chonnam.ac.kr}

\begin{abstract}
An explicit expression for the quadratic density-response function
of a many-electron system is obtained in the framework of the
time-dependent density-functional theory, in terms of the linear
and quadratic density-response functions of noninteracting
Kohn-Sham electrons and functional derivatives of the
time-dependent exchange-correlation potential. This is used to
evaluate the quadratic stopping power of a homogeneous electron
gas for slow ions, which is demonstrated to be equivalent to that
obtained up to second order in the ion charge in the framework of
a fully nonlinear scattering approach. Numerical calculations are
reported, thereby exploring the range of validity of quadratic-response theory.
\end{abstract}


\maketitle

\section{Introduction}
The inelastic interaction of charged particles with matter is one of
the fundamental problems of contemporary physics.
It encompasses such phenomena as the stopping power of solids for moving ions,
electron and positron energy-loss spectroscopy, inelastic low-energy electron
diffraction,
and hot-electron dynamics \cite{Echenique-90,Liebsch}.

A fruitful approach to the theoretical treatment of particle-solid interactions
has proven
to be the use of perturbation series expansions in powers of the
projectile-target Coulomb interaction.
For the description of many-electron targets, one typically introduces linear
and quadratic density-response
functions, which describe the electron density induced by external
perturbations.

While linear-response theory has proven successful in the
description of the interaction of fast projectiles with solids, in
the case of low projectile velocities and low electron densities a
nonlinear description becomes quantitatively
necessary \cite{Hu-88,Pitarke-93,Pitarke-95}. Besides, there exist
phenomena which cannot be explained in the framework of
linear-response theory, an example being the existing difference
between the scatterings of positively and negatively charged
particles \cite{Barkas-56,Moller-02}.

The random-phase approximation (RPA) has served as the natural
starting point for the calculation of both
linear \cite{Lindhard-54} and quadratic \cite{Pitarke-93,Pitarke-95}
density-response functions of the homogeneous electron gas.
However, exchange and correlation (xc) effects, which are absent
in the RPA, are known to be important for metallic electron
densities \cite{Gaztelurrutia-00}. The purpose of this paper is to
derive in the framework of the time-dependent density-functional
theory (TDDFT) \cite{Runge-84,Gross-96} an explicit expression for
the quadratic density-response function of a many-electron system,
which will then be used to evaluate the second-order energy loss
per unit path length of charged particles moving through solid
targets, i.e. the so-called stopping power of the target.

Another approach to evaluate the energy loss of slow ions moving in
a many-electron system is based on the ordinary formulation of
scattering theory. In this
approach \cite{Echenique-81,Echenique-86,Nagy-89}, the stopping power
for a heavy particle is determined in the low-velocity limit from
the knowledge of the scattering phase-shifts, which can be obtained
from a static nonlinearly screened potential by solving
self-consistently the Kohn-Sham equation of density-functional
theory (DFT) \cite{Kohn-65}. Since these nonperturbative calculations
include all orders in the projectile-target interaction, they
represent an important standard to investigate the range of validity
of perturbative expansions. Nonetheless, they have the limitation of
being restricted to low velocities ($v<<v_F$, $v_F$ being the Fermi
velocity) of recoilless probe particles moving in bulk materials.
\footnote{An extension of DFT-based potential-scattering calculations to
finite (although still small) projectile velocities has been reported
in Ref.~\cite{Salin-99}.
The interrelation of this approach and the quadratic response theory at finite
velocities is, however, beyond the scope of the present work.}

The interrelation of the perturbative-response and
nonperturbative-scattering approaches in their overlapping range of
applicability (low-velocity limit and small projectile charge) is
both an interesting and non-trivial problem. The starting points of
these two schemes are completely different and there are no grounds
to {\em a priori} assume equivalence between them. In this paper, we
demonstrate that in the low-velocity limit and to second order in
the external perturbation our quadratic-response formalism and the
scattering approach are equivalent, thereby extending the RPA-based
proof reported in Ref.~\cite{Hu-88} to the general case where the xc
effects are included.

This paper is organized as follows. In Sec.~\ref{main}, we derive in
the framework of TDDFT a formally exact explicit expression for the
quadratic density-response function of a many-electron system, in
terms of the noninteracting Kohn-Sham linear and quadratic
density-response functions and functional derivatives of the
time-dependent xc potential. In Sec.~\ref{stopping}, we derive basic
expressions for the stopping power of a uniform electron gas, in the
framework of both quadratic-response and nonperturbative-scattering
schemes. The results of numerical calculations are presented in
Sec.~\ref{num}. We use atomic units throughout, i.e.,
$e^2=\hbar=m_e=1$.

\section{Quadratic density response}
\label{main}

In the framework of TDDFT, Petersilka {\it et al.} \cite{Petersilka-96}
demonstrated that
within linear-response theory the electron density
$n_1({\bf r},t)$ induced in an arbitrary interacting many-electron system by
the time-dependent
external potential $\phi^{ext}({\bf r},t)$ coincides with the electron density
induced in
the corresponding system of noninteracting Kohn-Sham electrons by the
time-dependent effective potential
\begin{eqnarray}\label{eff1}
&&\phi_1^{eff}({\bf r},t)=\phi^{ext}({\bf r},t) +\int d {\bf
r}'\,v({\bf r},{\bf r}')\,n_1({\bf r}',t)\cr\cr &&+\int d {\bf
r}'\int dt'\,f_{xc}[n_0]({\bf r},t;{\bf r}',t')\,n_1({\bf r}',t'),
\end{eqnarray}
where $v({\bf r},{\bf r}')=1/|{\bf r}-{\bf r}'|$ is the bare Coulomb potential,
and $f_{xc}[n_0]({\bf r},t;{\bf r}',t')$ is the functional derivative of the
time-dependent
xc potential $V_{xc}[n]({\bf r},t)$ of TDDFT, to be evaluated at the
unperturbed
static electron density $n_0({\bf r})$:
\begin{equation}\label{fxc}
f_{xc}[n_0]({\bf r},t;{\bf r}',t')=\left.{\delta V_{xc}[n]({\bf
r},t)\over\delta n({\bf r}',t')}\right|_{n=n_0}.
\end{equation}

The linear-response scheme reported in Ref.~\cite{Petersilka-96} can
be extended to all orders in the external perturbation. This has
been carried out by Gross {\it et al.} \cite{Gross-96} in the general
case of spatially inhomogeneous electron systems, and
self-consistent integral equations for the quadratic and higher
order interacting density response functions have been obtained by
these authors. In the specific case of the uniform electron gas,
which we are here interested in, these equations can be easily
solved, to produce explicit interacting density response functions
in terms of their noninteracting counterparts and the functional
derivatives of the exchange-correlation potential. However, instead
of adopting the method of solution of the above mentioned integral
equations, we find it more instructive for our purposes, as well as
self-contained, to derive an explicit expression for the quadratic
density response function considering the uniform case from the very
beginning.

The electron density $n_1({\bf r},t)+n_2({\bf r},t)+\cdots$ induced
in an arbitrary many-electron system by the time-dependent external potential
$\phi^{ext}({\bf r},t)$
coincides with the electron density induced in the corresponding system of
noninteracting Kohn-Sham electrons
by the time-dependent effective potential $\phi_1^{eff}({\bf
r},t)+\phi_2^{eff}({\bf r},t)+\cdots$,
where $\phi_1^{eff}({\bf r},t)$ is given by Eq.~(\ref{eff1}) and
\begin{eqnarray}\label{eff2}
&&\phi_2^{eff}({\bf r},t)=\int d {\bf r}'\,v({\bf r},{\bf
r}')\,n_2({\bf r}',t)\cr\cr &&+\int d {\bf r}'\int
dt'\,f_{xc}[n_0]({\bf r},t;{\bf r}',t')\,n_2({\bf r}',t)\cr\cr &&+
\frac{1}{2} \int d {\bf r}'\int dt'\int d {\bf r}''\int
dt''\,g_{xc}[n_0]({\bf r},t;{\bf r}',t';{\bf r}'',t'')\cr\cr
&&\times n_1({\bf r}',t')\,n_1({\bf r}'',t''),
\end{eqnarray}
$g_{xc}[n_0]({\bf r},t;{\bf r}',t';{\bf r}'',t'')$ being the second functional
derivative of
the time-dependent xc potential $V_{xc}[n]({\bf r},t)$, to be evaluated at the
unperturbed static
electron density $n_0({\bf r})$:
\begin{equation}\label{gxc}
g_{xc}[n_0]({\bf r},t;{\bf r}',t';{\bf r}'',t'')= \left.{\delta^2
V_{xc}[n]({\bf r},t)\over\delta n({\bf r}',t') \delta n({\bf
r}'',t'')}\right|_{n=n_0} \hspace{-0.125 cm}.
\end{equation}

In the case of a homogeneous electron gas, there is translational
invariance in all directions. Hence, taking Fourier transforms
with respect to space and time, the {\it exact} momentum and
frequency dependent  induced electron densities, $n_1({\bf
q},\omega)$ and $n_2({\bf q},\omega)$, can be written as
\begin{equation}\label{dens1}
n_1(q)=\chi_1(q)\phi^{ext}(q)=\chi^0_1(q)\phi_1^{eff}(q)
\end{equation}
and
\begin{eqnarray}\label{dens2}
&&n_2(q)=\int d^4q_1\chi_2(q,q_1)\phi^{ext}(q_1)\phi^{ext}(q-q_1)=
\chi^0_1(q) \cr\cr &&\times \phi_2^{eff}(q) +\int
d^4q_1\chi^0_2(q,q_1)\phi_1^{eff}(q_1)\phi_1^{eff}(q-q_1),
\end{eqnarray}
where $q=({\bf q},\omega)$, $\phi_1^{eff}(q)$ and
$\phi_2^{eff}(q)$ are Fourier transforms of the time-dependent
effective potentials of Eqs.~(\ref{eff1}) and (\ref{eff2}), respectively,
$\phi^{ext}(q)$ is the Fourier transform of the external
potential, $\chi_1(q)$ and $\chi_2(q,q_1)$ denote the {\it exact}
linear and quadratic density-response functions of the interacting
electron system, and $\chi_1^0(q)$ and $\chi_2^0(q,q_1)$ represent
the corresponding density-response functions of noninteracting
Kohn-Sham electrons. Substituting Eqs.~(\ref{eff1}) and
(\ref{eff2}) into Eqs.~(\ref{dens1}) and (\ref{dens2}), one finds
\begin{equation}\label{chi1}
\chi_1(q)= \tilde\epsilon^{-1}(q) \, \chi_1^0(q)
\end{equation}
and
\begin{eqnarray}\label{chi2}
\chi_2(q,q_1)&=&\tilde\epsilon^{-1}(q)\chi_2^0(q,q_1)\tilde\epsilon^{-1}(q_1)
\tilde\epsilon^{-1}(q-q_1)\cr\cr
&+&\chi_1(q)g_{xc}(q,q_1)\chi_1(q_1) \chi_1(q-q_1)/2,
\end{eqnarray}
where $v(q)=4\pi/{\bf q}^2$ is the Fourier transform of the
Coulomb potential, $f_{xc}(q)$ and $g_{xc}(q,q_1)$ denote the
Fourier transforms of the xc kernels of Eqs.~(\ref{fxc}) and
(\ref{gxc}), and $\tilde\epsilon(q)$ is the test-charge--electron
dielectric function \cite{Kukkonen-79,Echenique-00}
\begin{equation}
\label{epst}
\tilde\epsilon(q)=1- \chi_1^0(q)\left[v(q)+f_{xc}(q)\right].
\end{equation}
The Fourier transforms of the linear and quadratic xc kernels in
Eqs.~(\ref{epst})
and (\ref{chi2}), respectively, are defined as
\begin{eqnarray*}
f_{xc}(q) &=& \int f_{xc}[n_0]({\bf r},t;{\bf r}',t')
e^{-i {\bf q}\cdot({\bf r}-{\bf r}') + i \omega (t-t')} d {\bf r} \, d t,
\cr\cr
g_{xc}(q,q_1) &=& \int g_{xc}[n_0]({\bf r},t;{\bf r}',t';{\bf r}'',t'')
e^{i [{\bf q}_1\cdot({\bf r}'-{\bf r}) -  \omega_1 (t'-t)]}\cr\cr
&\times&
e^{i [({\bf q}-{\bf q}_1)\cdot({\bf r}''-{\bf r})  - (\omega-\omega_1)
(t''-t)]} d {\bf r}' \, d t' \, d {\bf r}'' \, d t''.
\end{eqnarray*}

Equations~(\ref{chi1}) and (\ref{chi2}) generalize the {\it exact}
linear density response reported in Ref.~\cite{Petersilka-96} to the
realm of quadratic-response theory and the static quadratic density
response reported in Ref.~\cite{Louis-98} to the general case of a
time-dependent perturbation. In the so-called adiabatic LDA (ALDA),
which is only rigorous in the long-wavelength (${\bf q}\to 0$) and
static ($\omega\to 0$) limits, the xc kernels $f_{xc}(q)$ and
$g_{xc}(q,q_1)$ are simply the first and second derivatives with
respect to the unperturbed density of the static xc potential of a
uniform electron gas: $V'_{xc}(n_0)$ and $V''_{xc}(n_0)$. In the
RPA, the xc kernels $f_{xc}(q)$ and $g_{xc}(q,q_1)$ are set equal to
zero. Finally, we find our theory in agreement with that of
Ref.~\cite{Gross-96}, while the homogeneity of the system  we
consider enables us to obtain the explicit quadratic
density-response function of Eq.~(\ref{chi2}) instead of presenting
the results in the form of self-consistent integral equations.

\section{Stopping power}
\label{stopping}

There are two routes to describe the stopping power
of the homogeneous electron gas. One is based on a perturbative
expansion of the density response of the target (appropriate for
arbitrary projectile velocities) and the other on the knowledge of
the phase shifts of potential scattering of electrons by a statically
screened impurity (only valid for low projectile
velocities). We first consider these two alternative approaches,
then we focus on the overlapping range of their applicability, by considering
the low-velocity
limit of the quadratic-response formulation and a second-order
expansion of the transition-matrix elements of potential scattering.

\subsection{Quadratic density response}
\label{ALDA}

To third order in the projectile charge $Z_1$, the average energy
lost per unit length traveled by a recoilless probe particle
moving with velocity $\textbf{v}$ in a homogeneous electron gas,
i.e., the so-called stopping power of the target is obtained as
follows \cite{Gaztelurrutia-01,Nazarov-02}
\begin{eqnarray}
-\frac{d E}{d x}&=&-2\,\frac{Z_1^2}{\pi v }\int d {\bf
q}\,\frac{{\bf q}\cdot{\bf v}}{\textbf{q}^2}\,\left[ \frac{ {\rm
Im}\, \chi_1(q)}{\textbf{q}^2}\right.\cr\cr &+&\left.\frac{Z_1}{2
\pi^2} \int \frac{d {\bf q}_1}{\textbf{q}_1^2 |{\bf q} -{\bf
q}_1|^2} \, {\rm Im}\,\chi_2(q,q_1) \right], \label{E2}
\end{eqnarray}
which by virtue of Eqs.~(\ref{chi1}) and (\ref{chi2}) can be rigorously
expressed as
\begin{eqnarray}\label{stop2}
-\frac{d E}{d x}=-2\,\frac{Z_1^2}{\pi v } \, {\rm Im} \int d
{\bf q}\,\frac{{\bf q}\cdot{\bf v}}{\textbf{q}^2}\,\left\{
\frac{\chi^0_1(q)\, \tilde{\epsilon}^{-1}(q)}{\textbf{q}^2} + \frac{Z_1}{2
\pi^2} \int \frac{d
{\bf q}_1}{\textbf{q}_1^2 |{\bf q} - {\bf q}_1|^2} \right.\times\cr\cr
\hspace{-1 cm}
\left.
\tilde{\epsilon}^{-1}(q)\, \tilde{\epsilon}^{-1}(q_1) \,
\tilde{\epsilon}^{-1}(q-q_1)
\left[
\chi^0_2(q,q_1)+\chi^0_1(q)g_{xc}(q,q_1)
\chi^0_1(q_1)\chi^0_1(q-q_1)/2
\right]
\right\},
\end{eqnarray}
where now $q=({\bf q},{\bf q}\cdot{\bf v})$ and $q_1=({\bf q}_1,{\bf
q}_1\cdot{\bf v})$.

\subsubsection{Low-velocity limit}

At low frequencies we can write \cite{Lindhard-54}
\begin{eqnarray*}
{\rm Im}\, \chi^0_1(q) = \omega \, A_{\bf q}
\end{eqnarray*}
and \cite{Pitarke-95}
\begin{eqnarray*}
{\rm Im} \, \chi_2^0(q,q_1)= \omega \, B_{{\bf q},{\bf q}_1} +
\omega_1 \, B_{{\bf q}_1,{\bf q}} + (\omega - \omega_1) \, B_{{\bf
q}-{\bf q}_1,-{\bf q}_1},
\end{eqnarray*}
where
\begin{eqnarray*}
A_{\bf q} &=& -\frac{\Theta(2 k_F - q)}{2 \pi q },
\cr\cr
B_{{\bf q},{\bf q}_1}&=&2\,A_{\bf q}\,\frac{
\left(1-k_F^2/q_R^2\right)^{-1/2}}{|{\bf q}_1||{\bf q}-{\bf q}_1|}
\cr\cr &\times& {\rm sgn} \, (\cos \phi_{\bf q}) \Theta(q_R-k_F).
\end{eqnarray*}
Then in the ALDA one finds
\begin{eqnarray}
-\frac{d E}{d x}=-2\frac{Z_1^2}{\pi v } \int d {\bf q}
\frac{({\bf q}\cdot{\bf v})^2}{\textbf{q}^2} \left\{ \frac{A_{\bf
q}\, \tilde\epsilon_{\bf q}^{-2}}{\textbf{q}^2} + \frac{Z_1 }{
\pi^2 } \int \frac{d {\bf q}_1}{\textbf{q}_1^2 |{\bf q}_1 - {\bf
q}|^2} \right.\cr\cr
\hspace{-1.5 cm}
\times\left.\left[ \tilde{\epsilon}_{\bf
q}^{-1}\,B_{{\bf q},{\bf q}_1} \tilde{\epsilon}_{{\bf
q}_1}^{-1}\tilde{\epsilon}_{{\bf q}-{\bf q}_1}^{-1} +C_{\bf
q}\tilde\epsilon_{\bf q}^{-2}\chi^0_{2,({\bf q},{\bf q}_1)}
\tilde{\epsilon}^{-1}_{{\bf q}_1}\tilde{\epsilon}^{-1}_{{\bf
q}-{\bf q}_1} +V''_{xc}\,A_{\bf q}\tilde\epsilon^{-2}_{\bf
q}\,\chi_{1,{\bf q}_1}\,\chi_{1,{\bf q}-{\bf q}_1}/2, \right]
\right\}.\label{notsobig}
\end{eqnarray}
Here
\begin{eqnarray*}
C_{\bf q}=(v_{\bf q}+V'_{xc})\, A_{\bf q},
\end{eqnarray*}
$v_{\bf q}=4\pi/{\bf q}^2$ is the Fourier transform of the Coulomb potential,
$k_F$ is the Fermi momentum,
$\chi_{1,{\bf q}}$ and $\chi_{2,({\bf q},{\bf q}_1)}^0$ denote the static
($\omega=0$)
linear interacting and quadratic noninteracting density-response functions,
respectively, $\tilde\epsilon_{\bf q}$
is the static test-charge--electron dielectric function, $\Theta(x)$ is the
Heaviside step function, $q_R$
is the radius of the circle circumscribing the triangle formed by the vectors
${\bf q}$, ${\bf q}_1$,
and ${\bf q}-{\bf q}_1$, and $\phi_{\bf q}$ represents the angle facing ${\bf
q}$ in this triangle.
Evaluating some of the integrals in Eq.~(\ref{notsobig}), one finds
\begin{eqnarray}\label{res}
\hspace{-1.5 cm}
-\frac{1}{v}\,\frac{d E}{d x}=\frac{4Z_1^2}{3 \pi}
\int\limits_0^{2 k_F} d q \left\{ \frac{1}{q \,
\tilde{\epsilon}^2_{\bf q}} + \frac{2 Z_1 q}{\pi } \int_0^\infty
dq_1\int_{-1}^1d\mu \frac{1} {|\textbf{q}-\textbf{q}_1|^2
\tilde{\epsilon}_{\bf q} \tilde{\epsilon}_{{\bf q}_1}
\tilde{\epsilon}_{{\bf q}-{\bf q}_1}} \right.\\
\hspace{-1.5 cm}
\times\left.
\left[ \frac{(v_{\bf q} +V_{xc}') \chi^0_{2,({\bf q},{\bf q}_1)} +
V_{xc}'' \chi^0_{1,{\bf q}_1} \chi^0_{1,{\bf q}-{\bf q}_1}/2}
{\tilde\epsilon_{\bf q } }
 + \frac{2 \, \Theta(q_R-k_F)} {q_1|{\bf q}-{\bf q}_1|
\sqrt{1-k_F^2/q_R^2} } \,  {\rm sgn}(\cos\phi_{\bf q}) \right]
\right\}, \nonumber
\end{eqnarray}
where $q$ and $q_1$ now denote the magnitude of ${\bf q}$ and ${\bf
q}_1$, respectively, and $\mu=\cos\phi_{{\bf q}-{\bf q}_1}$. If we
put $V'_{xc}=V''_{xc}=0$ in Eq.~(\ref{res}), we retrieve the RPA
result of Ref.~\cite{Pitarke-95}. If we keep the actual value of
$V'_{xc}$ but still putting $V''_{xc}=0$, we reproduce the
calculations reported in Ref.~\cite{Gaztelurrutia-00}.

\subsection{Potential scattering}
\label{DFT}

In the low-velocity limit of a recoilless probe particle of
charge $Z_1$, the interaction between the Fermi gas and the probe
particle can be represented as the elastic scattering of
independent electrons by a Kohn-Sham effective static central
potential $V(r)$. Hence, the average energy loss per unit
path length of a recoilless charged particle moving with velocity ${\bf v}$
($v<<v_F$)
through a uniform electron gas of
density $n_0$ is given by the following expression:
\begin{equation}\label{scat1}
-\frac{d E}{d x}= n_0 v \, k_F \sigma_{tr}(k_F),
\end{equation}
where
\begin{equation}\label{scat2}
\sigma_{tr}(k)={16 \pi^5\over k^4}\int_0^{2k}dq \, q^3|T_{fi}|^2
\end{equation}
is the so-called transport cross section, $T_{fi}$ denoting the
transition-matrix element \cite{Taylor}:
\begin{equation}
T_{fi}=<\phi_{{\bf k}_f}|V|\psi_{{\bf k}_i}>.
\end{equation}
Here, $\phi_{\bf k}$ and $\psi_{\bf k}$ represent noninteracting
and interacting electron wave functions, respectively, ${\bf k}_i$
and ${\bf k}_f$ denote the electron momentum before and after the
collision, $k=|{\bf k}_i|=|{\bf k}_f|$ is the magnitude of the
electron momentum, ${\bf q}={\bf k}_f-{\bf k}_i$ is the momentum
transfer, and $V(r)$ is taken to be the Kohn-Sham effective
potential
\begin{equation}
V(r) = -\frac{Z_1}{r} + \int \frac{ n(r')}{|{\bf r}-{\bf r}'|} \, d {\bf r}' +
V_{xc}(r),
\label{VV}
\end{equation}
$n(r)$ being the electron density induced by the presence of the static probe
particle and
$V_{xc}(r)$ being the xc potential at point ${\bf r}$ of the inhomogeneous
electron system,
which in the LDA is simply the xc potential of a homogeneous electron gas with
electron density $n(r)$.
The well known expression of the transport cross-section in terms of the
phase-shifts $\delta_l(k)$ of the scattering problem in the spherically symmetric
potential
\begin{eqnarray}
\sigma_{tr}(k)=\frac{4\pi}{k^2} \sum\limits_{l=0}^{\infty} (l+1) \sin^2[\delta_l(k)-\delta_{l+1}(k)],
\label{trans}
\end{eqnarray}
greatly facilitates the numerical calculations, while the Friedel sum rule \cite{Kittel}
\begin{eqnarray}
Z_1=\frac{2}{\pi}\sum\limits_{l=0}^{\infty} (2 l+1) \delta_l(k_F)
\label{Fried}
\end{eqnarray}
is helpful in controlling self-consistency. Echenique {\it et al.}
\cite{Echenique-81,Echenique-86} and Nagy {\it et al.}
\cite{Nagy-89} evaluated the LDA Kohn-Sham effective potential by
solving self-consistently the Kohn-Sham equation of DFT, and then
computed the stopping power from Eqs.~(\ref{scat1}) and
(\ref{trans}).

We proceed by considering a second-order perturbative expansion of
the transition matrix $T_{fi}(q)$ entering Eq.~(\ref{scat2}),
which will then allow us to derive  $Z_1^2$- and $Z_1^3$-
contributions to the stopping power from Eq.~(\ref{scat1}).

The Born series for the transition matrix element $T_{fi}$ in powers of the
effective potential $V(r)$
[which can be expanded in powers of the bare interaction $-Z_1/r$:
$V_1(r)+V_2(r)\cdots$] is obtained as follows
\begin{equation}\label{tfi}
T_{fi}=\langle\phi_{{\bf k}_f}|\,V+VG_{k}^0V+\cdots\,|\phi_{{\bf k}_i}\rangle,
\end{equation}
where $G^0_{k}({\bf r},{\bf r}')$ is the noninteracting Green's function
\begin{eqnarray*}
G_k^0({\bf r},{\bf r}')=-{1\over 2\pi}\,{\rm e}^{ik|{\bf r}-{\bf
r}'|}/|{\bf r}-{\bf r}'|.
\end{eqnarray*}
Up to third order in the charge $Z_1$ of the probe particle, the
square of the transition matrix element of Eq.~(\ref{tfi}) yields
\begin{eqnarray}\label{exp}
&&|T_{fi}|^2=\left[V_1({\bf q})\right]^2+2\,V_1({\bf q})\cr\cr
&\times&\left[V_2({\bf q}) + P \int d {\bf q}_1\frac{V_1({\bf
q}_1)V_1({\bf q}-{\bf q}_1)} {{\bf k}_i^2/2-({\bf k}_i-{\bf
q}_1)^2/2 }\right],
\end{eqnarray}
where
\begin{equation}\label{v1}
V_1(\textbf{q})= -\frac{Z_1}{2 \pi^2 q^2}+(v_{\bf q}+V'_{xc}) \,
n_1(\textbf{q})
\end{equation}
and
\begin{eqnarray}\label{v2}
V_2(\textbf{q})&=&(v_{\bf q}+V'_{xc})\,n_2(\textbf{q})\cr\cr
&+&\frac{V''_{xc}}{2} \int d {\bf q}_1\,n_1({\bf q}_1)\, n_1({\bf
q}-{\bf q}_1),
\end{eqnarray}
$n_1({\bf q})$ and $n_2({\bf q})$ being the linear and quadratic
induced electron densities, and $P$ in Eq.~(\ref{exp}) denoting that
the principal value of the integral must be taken at the point where
the integrand is singular. It is interesting to notice that the
second-order ($Z_1^3$) contribution to $|T_{fi}|^2$ has two sources.
One is the first Born contribution to the quadratically screened
effective potential $V(r)$ and the other is the second Born
contribution to the linearly screened effective potential, as
pointed out in Ref.~\cite{Hu-88}. They have opposite signs and it is
the latter which dominates \cite{Hu-88,Pitarke-93,Pitarke-95}.

Substituting Eqs.~(\ref{dens1}) and (\ref{dens2}) into
Eqs.~(\ref{v1}) and (\ref{v2}) and then substituting the expansion
of Eq.~(\ref{exp}) into Eq.~(\ref{scat2}), one finds from
Eq.~(\ref{scat1}) the following expansion for the stopping power:
\begin{eqnarray}
-\frac{1}{v}\,\frac{dE}{d x}=\frac{4 }{3 \pi} \int\limits_0^{2
k_F} d q \left\{ \frac{Z_1^2}{q \tilde{\epsilon}^2_{\bf q}} +
\frac{Z_1^3 q}{\pi^2 } \int \frac{d {\bf q}_1}{q_1^2
|\textbf{q}-\textbf{q}_1|^2 \tilde{\epsilon}_{\bf q}
\tilde{\epsilon}_{{\bf q}_1} \tilde{\epsilon}_{{\bf q}-{\bf q}_1}
} \right.\cr\cr
\times\left.\left[ \frac{(v_{\bf q} +V_{xc}')
\chi^0_{2,({\bf q},{\bf q}_1)} + V_{xc}''\, \chi^0_{1,{\bf
q}_1}\,\chi^0_{1,{\bf q}-{\bf q}_1}/2} {\tilde{\epsilon}_{\bf q} }
+ \frac{2 }{k_F^2-({\bf k}_F-{\bf q}_1)^2} \right] \right\}.
\end{eqnarray}
Performing the integration over the angular variables of ${\bf
q}_1$, one readily reproduces Eq.~(\ref{res}), thereby proving the
equivalence between the quadratic-response and potential-scattering
schemes in the limit of low velocities of the probe particle. This
generalizes the RPA analysis reported in Ref.~\cite{Hu-88} to the
general situation where xc effects are taken into account.

\section{Results of numerical calculations}
\label{num}
\begin{figure}[h]
\includegraphics{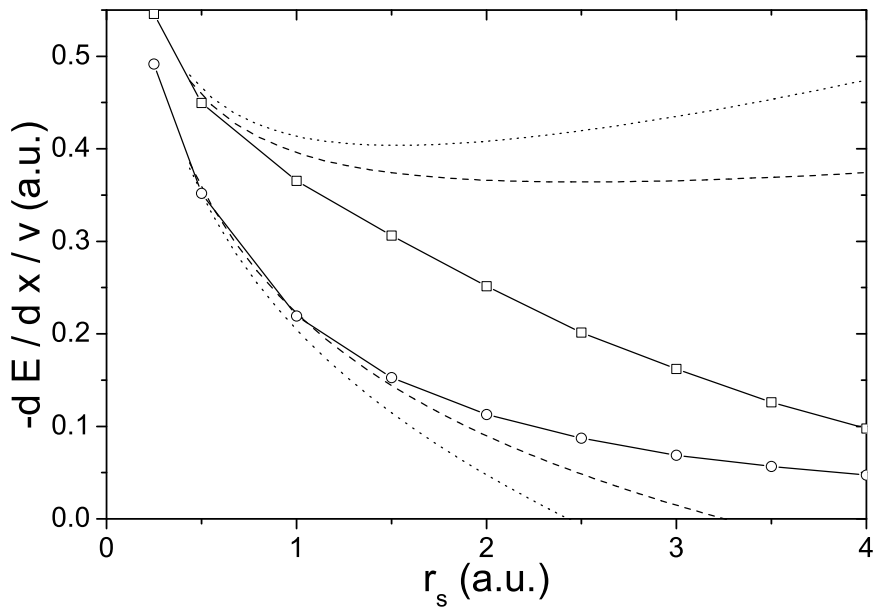}
\caption{\label{Z3} Stopping power of a uniform electron gas in
the low-velocity limit, divided by the projectile velocity
(friction coefficient), as a function of the electron-density
parameter $r_s$. The chained curves represent nonperturbative
potential-scattering
LDA calculations for protons (squares) and antiprotons (circles).
The dashed (dotted) line represents ALDA calculations to third order in the
projectile charge
$Z_1$ with (without) inclusion of the second derivative $V''_{xc}$ of the xc
potential.}
\end{figure}
\begin{figure}[h]
\includegraphics{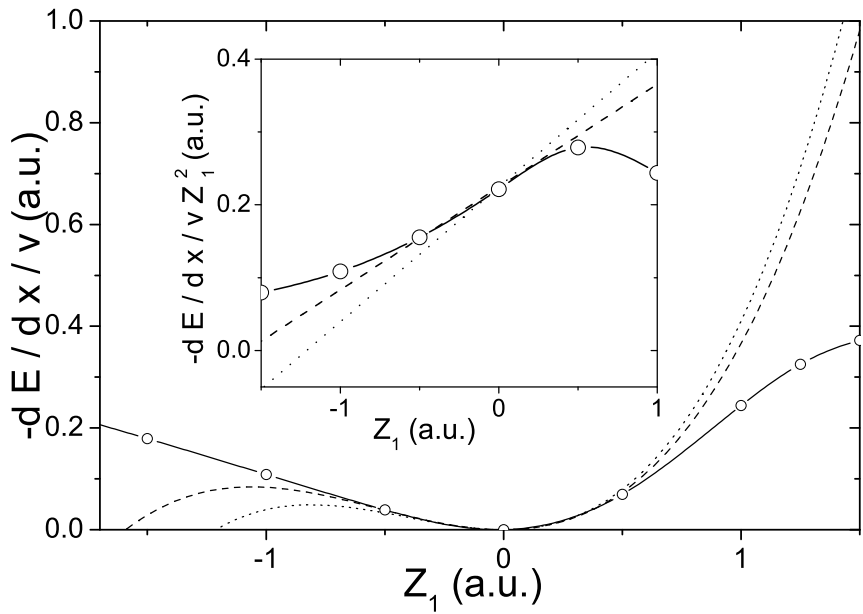}
\caption{\label{ZZ3} Stopping power of a uniform electron gas in
the low-velocity limit, divided by the projectile velocity
(friction coefficient), as a function of the projectile charge
$Z_1$ and for the electron-density parameter $r_s=2.07$
corresponding to the average electron density of valence electrons
in Al. The chained curve represents  nonperturbative
potential-scattering LDA calculation. The dashed (dotted) line
represents our ALDA calculations to third order in the projectile
charge $Z_1$ with (without) inclusion of the second derivative
$V''_{xc}$ of the xc potential. The inset shows the same plots
normalized to the square of the projectile's charge.}
\end{figure}
\begin{figure}[h]
\includegraphics{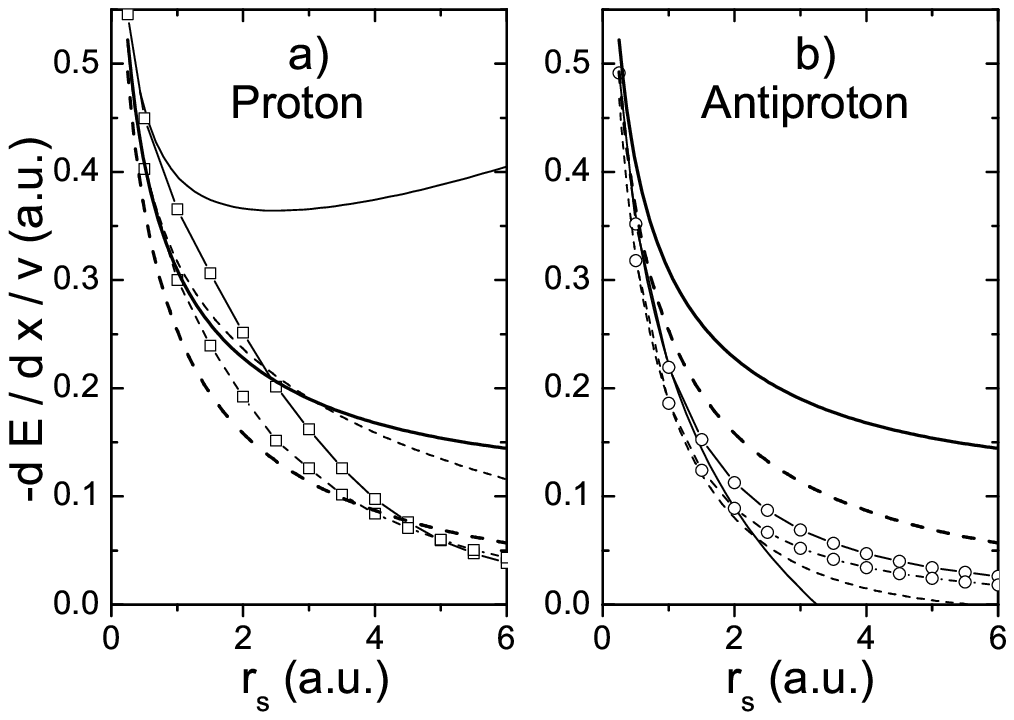}
\caption{\label{pl}
Stopping power of a uniform electron gas in
the low-velocity limit, divided by the projectile velocity
(friction coefficient) and as a function of the electron-density
parameter $r_s$, for protons (a) and antiprotons (b).
Solid (dashed) curves represent calculations including (omitting) exchange and
correlation. The chained curves represent the nonperturbative potential-scattering
calculations.
Bold (both solid and dashed) lines are linear ($Z_1^2$) contributions.
Thin (both solid and dashed) lines represent quadratic ($Z_1^3$)
calculations. }
\end{figure}

In Fig.~\ref{Z3} we plot the stopping power of a uniform electron
gas of density $n_0$ in the low-velocity limit, divided by the
projectile velocity (friction coefficient), for protons ($Z_1=1$)
and antiprotons ($Z_1=-1$) as a function of the electron-density
parameter $r_s=(3/4 \pi n_0)^{1/3}$. We evaluate both the
perturbative expansion of Eq.~(\ref{res}) and the non-perturbative
formula\footnote{For the non-perturbative calculation, we have used
the conventional scheme \cite{Echenique-81,Echenique-86} of
iterative solution of Kohn-Sham equations with the potential of
Eq.~(\ref{VV}) and calculation of the stopping power from
Eqs.~(\ref{scat1}) and (\ref{trans}) upon the achievement of
convergence. The fulfillment of the Friedel sum rule of
Eq.~(\ref{Fried}) has been monitored, the error in which has not
been greater than 0.02 electrons in all the calculations.} by using
the Perdew-Zunger \cite{Perdew-81} parametrization of the xc
potential of a uniform electron gas.  Our non-perturbative
calculations reproduce those reported in Refs.~\cite{Echenique-86}
and \cite{Nagy-89} for protons and antiprotons, respectively.
Perturbative and nonperturbative calculations are also plotted in
Fig.~\ref{ZZ3}, but now for the electron-density parameter
$r_s=2.07$ corresponding to valence electrons in Al and as a
function of the projectile charge $Z_1$.

Also plotted in Figs.~\ref{Z3} and \ref{ZZ3} are perturbative
calculations with no inclusion of the second derivative of the xc
potential $V_{xc}$, as reported in Ref.~\cite{Gaztelurrutia-00},
showing that the inclusion of this term brings the perturbative
calculations very close to the full  nonlinear
calculation in the range of high electron densities (small $r_s$)
and small projectile charges. Figure \ref{Z3} shows that the
quadratic (perturbative) stopping power for antiprotons is extremely
accurate for all electron densities with $r_s\leq 2$.
Figure~\ref{ZZ3} shows that in the case of Al target ($r_s=2.07$)
and negative projectile charges
the quadratic stopping power is accurate for the antiproton charge ($Z_1=-1$)
and above, but it is only
accurate for small positive values of the projectile charge
($Z_1\leq 0.5$)\footnote{Since $Z_1$ is the bare nucleus charge of the projectile,
for non-integer $Z_1$ the results should be considered as mathematical.}.
This is due to the presence of the truly bound
electronic states and the behavior of
resonances in the case of a positive probe
particle, which are only included in a fully nonlinear scheme.

To elucidate the role of resonances, we have performed the numerical analysis of the
phase-shifts $\delta_l(k)$
of the scattering in the self-consistent potential (\ref{VV}).
At a given $r_s$ with growing $Z_1$ a resonance,
which always exists in the continuum spectrum,
moves to lower energy and grows both sharper and more intense, which results in a
stronger variation of the phase-shifts. Since the density of states is proportional to the derivative of the phase-shifts
$\delta'_l(k)$, the low-lying continuum states get filled preferentially resulting,
similarly to the occupation of the bound states,
in the more efficient screening of the ion charge and eventually in
the decrease of the stopping power even before the formation of the bound states. On the other hand,
at smaller values of $Z_1$
the existence of weak broad resonances at high energies does not affect the applicability
of the quadratic theory.

In order to investigate the interplay between high-order
interactions and xc effects, we have plotted in Fig.~\ref{pl} the
results of linear ($Z_1^2$), quadratic ($Z_1^3$), and fully
nonlinear potential-scattering calculations of the stopping power of
slow protons and antiprotons, both in the absence and in the
presence of xc effects. This figure shows that: (i) The impact of xc
effects is considerably larger within linear and quadratic response
theory than in the more realistic nonperturbative
potential-scattering approach, especially so in the case of protons,
which indicates that there must be a large degree of cancelation
between first, second and higher-order xc effects. (ii) At high
electron densities ($r_s\to 0$), the inclusion of xc effects brings
the quadratic-response calculations (thin solid lines of
Fig.~\ref{pl}) into nice agreement with their DFT based
potential-scattering counterparts (chained solid lines). However, xc
effects decrease the radius of convergence of the asymptotic
perturbative expansion; while an unphysical negative stopping power
is obtained within RPA for antiprotons at $r_s>5.5$, this unphysical
behavior is obtained in the presence of xc effects at $r_s>3.2$.
(iii) The performance of the perturbative expansion is considerably
better for antiprotons than for protons. This can be attributed to
the existence of electronic bound states and the behavior of
resonances around a moving proton \cite{Echenique-86},
which are out of the reach of the perturbative description.
(iv) The performance of the quadratic response theory
for the stopping power is considerably better than in the case of
the electron density induced at the position of the projectile
\cite{Bergara-97,Arbo-01}, which is a highly nonlinear magnitude.
This is due to the fact that the stopping power involves an
integration of the induced density over the whole space, as
discussed in Ref.~\cite{Bergara-97}.

Finally, we note that apart from the obvious usefulness of
quadratic-response calculations in situations where the
interaction can be considered to be weak, it has been recently
shown that perturbative calculations can be successfully used as
input in a variational theory of charged particles interacting
with a many-body system. Recent investigations have shown that
this new variational theory brings the RPA quadratic stopping
power for slow antiprotons into nice agreement with the
corresponding nonperturbative potential-scattering calculations
for all electron densities \cite{NNP}.

\section{Summary and conclusions}\label{conc}

We have derived an explicit expression for the quadratic
density-response function of a many-electron system in the framework
of TDDFT, in terms of the linear and quadratic density-response
functions of noninteracting Kohn-Sham electrons and functional
derivatives of the time-dependent xc potential. This expression
generalizes the rigorous linear density-response function reported
in Ref.~\cite{Petersilka-96} to the realm of quadratic-response
theory, and they satisfy the self-consistent integral equations of
Gross {\it et al.} \cite{Gross-96}, valid for arbitrary inhomogeneous electron
system.

The exact expression for the quadratic density-response function has
been used to obtain the stopping power of a uniform electron gas to
second order in the projectile charge $Z_1$, which in the
low-velocity limit and within the adiabatic LDA is demonstrated to be
equivalent to that obtained up to third order in $Z_1$ in the
framework of a fully nonlinear LDA potential-scattering approach.
This generalizes the RPA analysis reported in Ref.~\cite{Hu-88} to
the general situation where xc effects are taken into account.

We have carried out LDA numerical calculations of the stopping power
of a uniform electron gas for slow positively and negatively charged
ions, as a function of both the electron-density parameter and the
projectile charge. We find that quadratic-response theory yields a
stopping power that is in excellent agreement with the
nonperturbative  stopping power in the range of
high electron densities and small projectile charges. The
quadratic-response (perturbative) stopping power for antiprotons is
found to be extremely accurate for all electron densities higher
than the electron density of valence electrons in Al. In the case of
Al, quadratic-response theory is found to yield accurate results for
small negative projectile charges up to the antiproton charge, but a
fully nonlinear scheme is required to account for the energy loss of
slow protons.

Although our equation (\ref{stop2}) for the stopping power is exact to the
$Z_1^3$ order,
in the numerical calculations we have utilized the local and adiabatic
approximation for the linear
and quadratic exchange-correlation kernels, which is consistent with the
available fully nonlinear calculations
within the potential scattering method. To study the role of the non-locality
(wave-vector dependence of the exchange-correlation potential) and
non-adiabaticity (its frequency dependence)
in the nonlinear
theory of stopping-power is, however, a challenging task, and this work is now
in progress
\cite{Nazarov-04-u}.

\ack

V.U.N. and C.S.K. acknowledge support by the Korea Research
Foundation through Grant No. KRF-2003-015-C00214. V.U.N.
acknowledges the Visiting Professorship at the Institute for Solid
State Physics of the University of Tokyo, a partial support by the
Grant-in-Aid for Scientific Research from the Ministry of Education,
Science, Sports, and Culture of Japan, and the hospitality of the
Donostia International Physics Center. J.M.P. acknowledges partial
support by the University of the Basque Country, the Basque
Hezkuntza, Unibertsitate eta Ikerketa Saila, and the Spanish
Ministerio de Ciencia y Tecnolog\'\i a.

\end{document}